\documentclass[%
superscriptaddress,
reprint,
amsmath,amssymb,
aps,
pra,
]{revtex4-1}

\usepackage{graphicx}% Include figure files
\usepackage[percent]{overpic}
\usepackage{dcolumn}% Align table columns on decimal point
\usepackage{bm}% bold math

\usepackage{hyperref}% add hypertext capabilities
\usepackage{color}
\definecolor{rltred}{rgb}{0.75,0,0}
\definecolor{rltgreen}{rgb}{0,0.6,0}
\definecolor{rltblue}{rgb}{0.3,0.3,1}

\usepackage{placeins}

\hypersetup{colorlinks,%
    hypertexnames=true,%
    linkcolor=rltred,%
    citecolor=rltgreen,%
    linktocpage=true}
  
\usepackage{physics}
\usepackage[utf8]{inputenc}
\usepackage[T1]{fontenc}
\begin{document}
\title{Chaos-induced depletion of a Bose-Einstein condensate}

\author{Ralf Wanzenböck}
\affiliation{Institute for Theoretical Physics, Vienna University of Technology,
	Wiedner Hauptstra\ss e 8-10/136, 1040 Vienna, Austria, EU}

\author{Stefan Donsa}
\affiliation{Institute for Theoretical Physics, Vienna University of Technology,
	Wiedner Hauptstra\ss e 8-10/136, 1040 Vienna, Austria, EU}

\author{Harald Hofstätter}
\affiliation{Institute for Theoretical Physics, Vienna University of Technology,
	Wiedner Hauptstra\ss e 8-10/136, 1040 Vienna, Austria, EU}

\author{Othmar Koch}
\affiliation{Faculty of Mathematics, University of Vienna, Oskar-Morgenstern-Platz 1, 1090 Vienna, Austria, EU}

\author{Peter Schlagheck}
\affiliation{Département de Physique, University of Liège, 4000 Liège, Belgium, EU}

\author{Iva B\v rezinov\'a}
\email{iva.brezinova@tuwien.ac.at}
\affiliation{Institute for Theoretical Physics, Vienna University of Technology,
    Wiedner Hauptstra\ss e 8-10/136, 1040 Vienna, Austria, EU}

\date{\today}

\begin{abstract}
The mean-field limit of a bosonic quantum many-body system is described by (mostly) non-linear equations of motion which may exhibit chaos very much in the spirit of classical particle chaos, i.e.~by an exponential separation of trajectories in Hilbert space with a rate given by a positive Lyapunov exponent $\lambda$. The question now is whether $\lambda$ imprints itself onto measurable observables of the underlying quantum many-body system even at finite particle numbers. Using a Bose-Einstein condensate expanding in a shallow potential landscape as a paradigmatic example for a bosonic quantum many-body system, we show, that the number of non-condensed particles is subject to an exponentially fast increase, i.e.~depletion. Furthermore, we show that the rate of exponential depletion is given by the Lyapunov exponent associated with the chaotic mean-field dynamics. Finally, we demonstrate that this chaos-induced depletion is accessible experimentally through the visibility of interference fringes in the total density after time of flight, thus opening the possibility to measure $\lambda$, and with it, the interplay between chaos and non-equilibrium quantum matter, in a real experiment. 
\end{abstract}
%%%%%%%%%%%%%%%%%%%%%%%%%%%%%%%%%%%%%%%%%%%%%%%%%%%%%%%%%%%%%%%%%%%%%%%%%%

\maketitle

\begin{figure}[t]
	\includegraphics[width=\columnwidth]{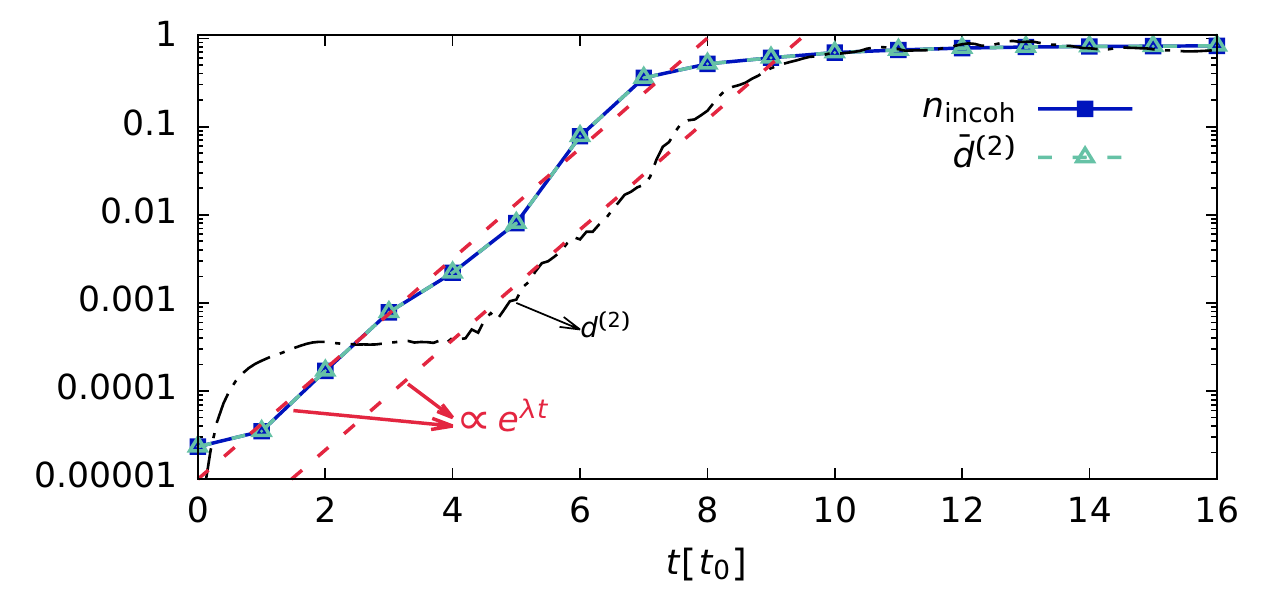}
	\caption{$d^{(2)}$ (Eq.~\ref{eq:d2}) for two initial conditions obtained by linear distortion of the mean-field ground state (black dashed-dotted line), distance function $\bar d^{(2)}$ (Eq.~\ref{eq:bar_d2}) averaged over the stochastic ensemble (triangles), and fraction of incoherent particles $n_\text{incoh}$ (squares). Data shown at integer values of $t$ (except for $d^{(2)}$), the lines serve as guides for the eye. The red dashed lines mark an exponential increase with $\lambda=1.44t_0^{-1}$. Number of particles is $N=1.2\times 10^6$, periodic potential $V(x)=V_P\cos{(k_Px)}$ used with $V_P=0.3e$, $k_P=\pi/3\xi$ and $\xi$ the healing length.}
	\label{fig:N6}
\end{figure}

%%%%%%%%%%%%%%%%%%%%%%%%%%%%%%%%%%%%%%%%%%%%%%%%%%%%%%%%%%%%%%%%%%%%%%%%%%
%\section{Introduction}\label{sec:intro}
%%%%%%%%%%%%%%%%%%%%%%%%%%%%%%%%%%%%%%%%%%%%%%%%%%%%%%%%%%%%%%%%%%%%%%%%%%
Non-equilibrium quantum many-body systems are nowadays routinely probed in experiments with ultracold atoms with unprecedented control over their parameters such as particle number, interaction strength, and external potentials. A plethora of non-equilibrium systems has been realized to address a wide variety of physics questions, including Anderson localization of Bose-Einstein condensates (BECs) \cite{clement_suppression_2005, fort_effect_2005, schulte_routes_2005, billy_direct_2008, roati_anderson_2008}, many-body localization in disordered lattices \cite{schreiber_observation_2015, lukin_probing_2019}, pre-thermalization of one-dimensional (1D) BECs \cite{gring_relaxation_2012}, and quench-dynamics of spin-model systems \cite{bernien_probing_2017}, to name just a few.\\ 
At the same time, the theoretical understanding of non-equilibrium quantum many-body systems still lags behind that obtained for stationary or equilibrium systems. In particular, the role of ``classical'' chaos on non-equilibrium quantum many-body systems is currently subject of intense scrutiny, see e.g.~\cite{ho_periodic_2019, hallam_lyapunov_2019, lewis-swan_unifying_2019, xu_does_2020}. For a bosonic quantum many-body system, the mean-field limit can be viewed as the classical limit in which the particle creation and annihilation operators lose their quantum properties and start to act as classical fields. The mean-field limit, typically involves non-linear (partial) differential equations, and can exhibit chaos with exponential separation in Hilbert space characterized by a positive Lyapunov exponent $\lambda$ \cite{brezinova_wave_2011, cassidy_threshold_2009}. One of the open questions is, whether wave chaos, more specifically the positive $\lambda$, imprints itself onto the dynamics of a quantum many-body system even at finite particle numbers and how such an imprint could be measured. Recently, out-of-time-order correlators have been suggested as suitable probes for a positive $\lambda$ (see e.g.~\cite{Kitaev2014,sekino_fast_2008,shenker_black_2014,maldacena_bound_2016,lewis-swan_unifying_2019, xu_does_2020}). \\
In this paper, we find an imprint of chaos on a different observable within a paradigmatic bosonic system: A quasi 1D BEC initially trapped harmonically and then released to expand in a shallow disordered or periodic potential. We show that the fraction of non-condensed particles increases exponentially over time and that the associated rate is given by the Lyapunov exponent $\lambda$ obtained from mean-field chaos. The depletion occurs on time scales during which most of the initial interaction energy is converted into kinetic energy, and comes to a halt at times close to the so-called scrambling time (or Ehrenfest time) \cite{rammensee_many-body_2018, tomsovic_post-ehrenfest_2018, maldacena_bound_2016}. We observe chaos-induced depletion both in shallow disordered as well as periodic potentials showing that the effect is quite general and does not rely on the intrinsic randomness of a disordered landscape.\\
Finally, we demonstrate that the condensate depletion and thus the Lyapunov exponent $\lambda$ is accessible experimentally through the analysis of fluctuations of the total particle density in momentum space. While the condensed part is coherent and leads to interference fringes in the total density, the non-condensed part is incoherent and piles up over time as a non-fluctuating background. Analyzing the interference fringes after time of flight would thus allow to extract experimentally the fraction of non-condensed particles as a function of time and compare to the theoretically obtained $\lambda$. Condensate depletion thus offers itself as an experimentally accessible probe to investigate the role of (mean-field or classical) chaos in non-equilibrium quantum matter.\\
%%%%%%%%%%%%%%%%%%%%%%%%%%%%
%System under investigation
%%%%%%%%%%%%%%%%%%%%%%%%%%%%
While our findings are generally applicable to bosonic systems that exhibit mean-field chaos, we pick one specific system already realized experimentally \cite{billy_direct_2008} to obtain numerical results. The initially harmonically trapped quasi 1D BEC of $N$ $^{87}$Rb atoms is released at $t=0$ to expand in a shallow potential. As units we use $\hbar=m=\omega_0=1$, with $\omega_0$ being the frequency of the initial longitudinal harmonic trap which amounts to a time unit of $t_0\approx30$ms and a space unit of $x_0\approx4.6\mu$m. For the number of atoms, we take $N=1.2\times 10^4$ following \cite{billy_direct_2008}, as well as larger values, i.e.~$N=1.2\times10^5$ and $N=1.2\times10^6$ to investigate the effect of varying $N$.\\
Describing the quasi-1D system on a mean-field level the Gross-Pitaevskii equation (GPE) takes the form
\begin{equation}
i\frac{\partial \psi(x,t)}{\partial t} = 
\left(-\frac{1}{2}\frac{\partial^2}{\partial x^2}
+V(x)+g|\psi(x,t)|^2\right)\psi(x,t),
\label{eq:gpe}
\end{equation}
where the nonlinearity is $g\approx400$ with the above parameters and the normalization $\int dx|\psi(x,t)|^2=1$. The potential $V(x)$ corresponds to the harmonic potential at $t=0$, and to the periodic or disordered speckle potential at $t>0$ with amplitude much smaller than the mean energy per particle $e$. This system exhibits chaos on the mean-field level \cite{brezinova_wave_2011}: Two wave functions, $\psi_a(x,0)$ and $\psi_{b}(x,0)$, respectively, initially very close to each other in Hilbert space as measured by a distance norm, separate exponentially in time until quasi-orthogonality is reached, see Fig.~\ref{fig:N6}.
The rate of the exponential growth is given by the Lyapunov exponent $\lambda$. For the distance norm we take
\begin{align}
d^{(2)}_{a,b}(t) &= \frac{1}{2}\int dx |\psi_a(x,t)-\psi_{b}(x,t)|^2 .
\label{eq:d2}
\end{align}
The Lyapunov exponent $\lambda$ shows systematic trends as a function of the parameters of the system: It vanishes in absence of inter-particle interactions for arbitrary potentials, as well as in presence of inter-particle interactions in free space (i.e.~without any potential). At fixed period of the periodic potential $k_P$, or fixed correlation length $\sigma$ of the speckle potential, it increases both with nonlinearity and the potential amplitude \cite{brezinova_wave_2011}, see the supplemental material (SM).\\
To find imprints of chaos on measurable observables of the quantum many-body system with finite $N$ a theory beyond mean-field has to be applied. The multi-configurational time-dependent Hartree method for bosons (see, e.g.~\cite{alon_multiconfigurational_2008,lode_colloquium_2020}), while being in principle exact for a sufficient number of orbitals, suffers from the exponentially growing configuration space. For the particle numbers considered, only two orbitals can be afforded numerically \cite{brezinova_wave_2012}. As more than two orbitals are populated during the propagation, the MCTHB method entails a large and not easy to quantify error. We, therefore, resort to the truncated Wigner approximation (TWA), see e.g.~\cite{Steel1998,Sinatra2002,blakie_dynamics_2008,dujardin_breakdown_2016} which employs the Wigner representation $W$ for (in general) a many-body density matrix $\hat \rho$
\begin{align}
W(&\psi_1,\ldots,\psi_M,\psi_1^* \ldots,\psi_M^*) = \frac{1}{\pi^{2M}}\nonumber \\
&\times \int dz^{2M}\Tr\left[\hat \rho e^{i\sum_j\left( z_j^*\hat\psi_j^\dagger
+iz_j\hat{\psi_j}\right)}
\right]
e^{-i\sum_j\left(z_j^*\psi_j^*-iz_j\psi_j\right)}.
\end{align}
$W$ can be viewed as a phase-space representation of the quantum many-body state. $M$ is the total number of modes in which particles can be created or annihilated and $\hat \psi_j^\dagger$ and $\hat \psi_j$ are the corresponding creation and annihilation operators, respectively. In general, particles can be created or annihilated in an arbitrary single particle mode denoted by $j$. We choose $j$ to represent a specific point in space assuming for simplicity an equidistant spatial discretization. We have made sure, however, that the spatial grid is fine enough, i.e.~the distance between grid points $dx<1/k_\text{max}$ with $k_\text{max}$ being the largest relevant momentum in the system, such that we are still in the continuum limit.\\ 
Having $W$ as a function of time at disposal would allow to evaluate all expectation values of symmetrized products of creation and annihilation operators. The exact equation of motion for $W$ can be obtained using von Neumann's equation of motion for $\hat\rho$. However, it proves to be intractable, such that approximations have to be invoked. Within the TWA \cite{Steel1998, Sinatra2002, blakie_dynamics_2008}, the 
time evolution of $W$ is sampled stochastically with an ensemble of trajectories obeying the GPE, Eq.~\ref{eq:gpe}. (The only modification comes from the fact that we have to discretize space such that the second derivative in Eq.~\ref{eq:gpe} has to be replaced by its second-order finite difference approximation.) It has been shown \cite{schlagheck_enhancement_2019, tomsovic_post-ehrenfest_2018, dujardin_describing_2015} that this approximation amounts to neglecting non-classical trajectories as well as interferences between distinct trajectories in many-body Hilbert space. The question then arises at which times do these neglected effects start to play a role and become non-negligible. For single- or few-particle systems, sampling the time evolution with classical trajectories is accurate up to the point where an initially maximally localized state has spread over the whole system. This time is called the Ehrenfest time $\tau_E$ \cite{ehrenfest_bemerkung_1927} which is, in presence of classical chaos, inversely proportional to $\lambda$ and grows logarithmically with $1/\hbar$. This concept can be extended into the many-body regime for bosonic systems with $\hbar$ being replaced by the effective Planck constant $\hbar_\text{eff}\simeq 1/N$. Following the lines of \cite{rammensee_many-body_2018, tomsovic_post-ehrenfest_2018} we thus assume that our results are accurate up to the time $\tau_E=\frac{1}{\lambda}\log{N}$.\\
The initial conditions within the stochastic ensemble of trajectories are constructed such as to correctly sample the phase-space distribution of the underlying initial quantum state, which in our case is a BEC at zero temperature. The stochasticity of the ensemble comes solely from the sampling of this initial state since Eq.~\ref{eq:gpe} is completely deterministic. We follow \cite{Sinatra2002, blakie_dynamics_2008, Steel1998} and construct the initial wave functions by adding to the mean-field ground state in the harmonic trap vacuum fluctuations in form of Gaussian noise (see the SM).\\
The most relevant observables in our case will be the coherent part of the particle density given by $\rho_\text{coh}(x_j,t) = |\langle \hat \psi_j(t) \rangle|^2$, as well as the one-particle reduced density matrix (1RDM) $D_{ij}(t)=\langle \hat\psi_i^\dagger(t)\hat\psi_j(t)\rangle$. The term ``coherent" in defining $\rho_\text{coh}(x_j,t)$ points to the fact that only a macroscopically occupied state with a spatially non-random phase will survive the averaging. $\rho_\text{coh}(x_j,t)$ can therefore be associated with the density of condensed particles. Alternatively \cite{penrose_bose-einstein_1956,leggett_bose-einstein_2001}, the condensate state is defined through a macroscopic occupation of one eigenstate of the 1RDM. We show in the SM that these two definitions of the condensate give practically identical results for the depletion over time such that we use throughout the remainder of the paper the term coherent synonymously to condensed.\\
Within the stochastic ensemble of trajectories, expectation values can be calculated as
%
%\begin{eqnarray}
$\langle \hat\psi_j(t)\rangle = \frac{1}{N_s}\sum_{s=1}^{N_s}\psi_s(x_j,t)
$,
%\label{eq:av_psi}
%\end{eqnarray}
%
with $N_s$ ($\gg1$) being the number of Gross-Pitaevskii trajectories $\psi_{s}(x_j,t)$ within the ensemble. To calculate the 1RDM, one has to rewrite $\langle \hat\psi_i^\dagger\hat\psi_j + \hat \psi_j\hat\psi_i^\dagger\rangle = 
2\langle \hat\psi_i^\dagger\hat\psi_j \rangle + \frac{1}{Ndx}\delta_{ij}$ using the commutator relation $[\hat\psi_j,\hat\psi_i^\dagger] = \frac{1}{Ndx}\delta_{ij}$.
The term $\delta_{ij}/dx$ is the discrete version of the $\delta$-function for a continuous system, and 
the factor $1/N$ comes from our normalization of the wave functions of Eq.~\ref{eq:gpe} to one, or equivalently, the creation and annihilation operators to $1/N$.
The 1RDM is then given by
%\begin{eqnarray}
$D_{ij}(t) = \frac{1}{N_s}\sum_{s=1}^{N_s}\psi^{*}_s(x_i,t)\psi_s(x_j,t) - \frac{1}{2Ndx}\delta_{ij}
$,
%	\label{eq:1rdm}
%\end{eqnarray}
and the total particle density is $\rho_\text{total}(x_j,t) = D_{jj}(t)$. The fraction of coherent particles is determined by 
$n_\text{coh}(t)=\sum_j dx\rho_\text{coh}(x_j,t)$. Accordingly, the fraction of incoherent particles is $n_\text{incoh}(t) = 1-n_\text{coh}(t)$. The crucial observation now is that
\begin{align}
n_\text{incoh}(t) = \frac{1}{N_s^2}\sum_{s,r}d^{(2)}_{s,r}(t)
-\frac{L}{2Ndx} = \bar d^{(2)}(t),
\label{eq:bar_d2}
\end{align}
which we obtain using Eq.~\ref{eq:d2}, taking into account that the norm of the wave functions within the ensemble is $1/N_s\sum_s\sum_j dx |\psi_s(x_j)|^2 = 1 + L/2Ndx$ with $L$ the length of the system. (Note that the term $L/dx$ counts the number of single-particle modes to which vacuum fluctuations have been added.) For a detailed derivation, see the SM. The right-hand side of Eq.~\ref{eq:bar_d2} is (apart from a constant term) the arithmetic mean over the distance function between all pairs of mean-field trajectories, and we denote it with $\bar d^{(2)}(t)$.\\
We have explicitly verified the equality of Eq.~\ref{eq:bar_d2} numerically by independently calculating the arithmetic mean of the distance function, $\bar d^{(2)}(t)$, and comparing it to $n_\text{incoh}(t)$, see Fig.~\ref{fig:N6}.
\begin{figure}[t]
	\includegraphics[width=\columnwidth]{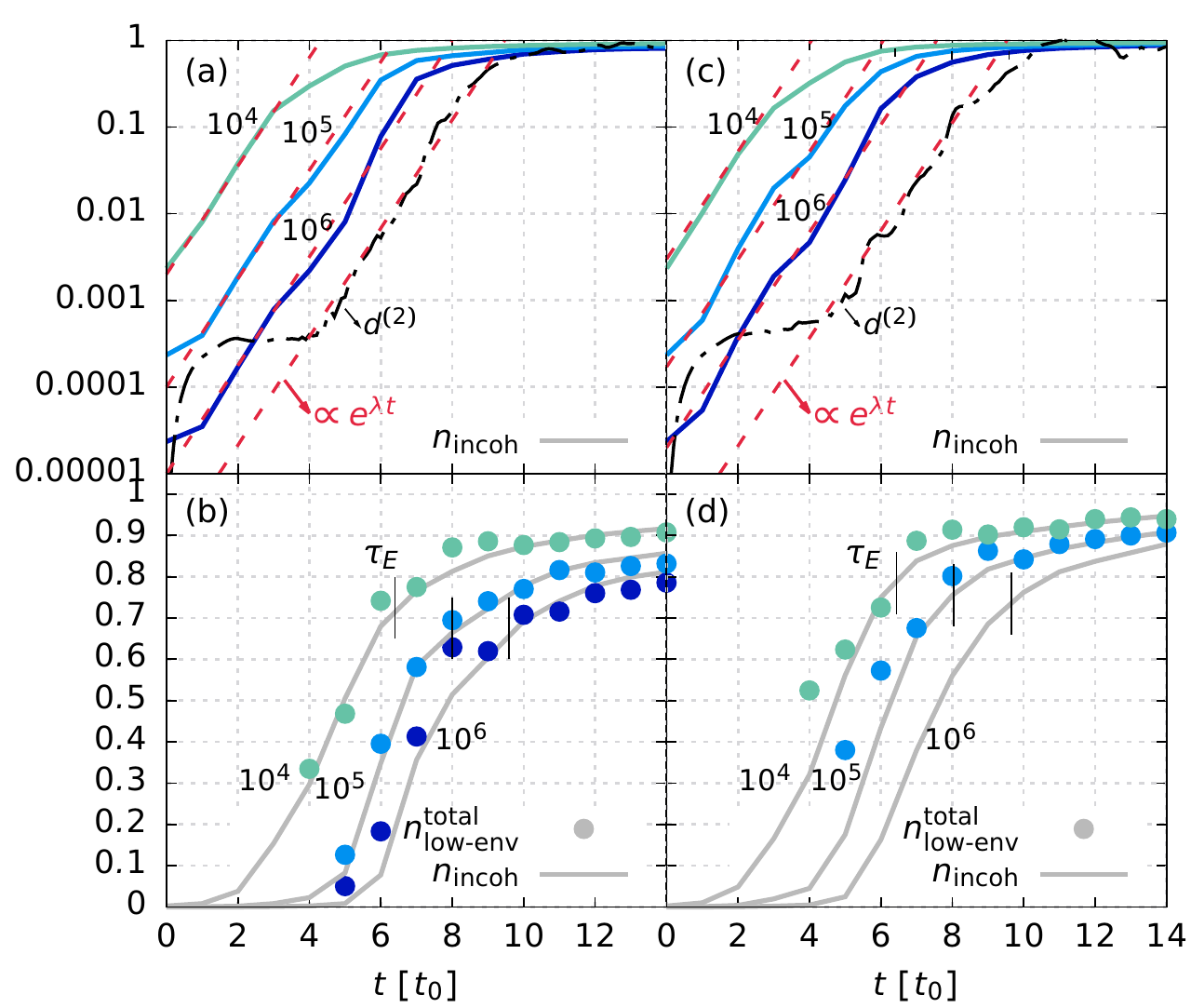}
	\caption{Left column (a) and (b) for the periodic potential with $V_P=0.3e$ and $k_P=\pi/3\xi$: (a) Fraction of incoherent particles $n_\text{incoh}$ for different particle numbers $N$ ($N$ being $1.2$ times the number near each curve), and $d^{(2)}$ for the linearly distorted initial conditions. Red dashed lines correspond to an exponential increase with $\lambda=1.44 t_0^{-1}$. (b) Linear plot of (a) including $n^\text{total}_\text{low-env}$ extracted from the total density only, see Fig.~\ref{fig:incoh_den}. The Ehrenfest time $\tau_E=1/\lambda \ln{N}$ is marked for each curve. Right column (c) and (d) same as left column but for one realization of speckle disorder with $V_D=0.3e$ and correlation length $\sigma =0.57\xi$. In (c) the Lyapunov exponent is $\lambda =1.43 t_0^{-1}$. Data is shown at integer values of $t$ (except for $d^{(2)}$), the lines serve as guides for the eye.}
	\label{fig:N_numb}
\end{figure}
\begin{figure}[t]
	\includegraphics[width=\columnwidth]{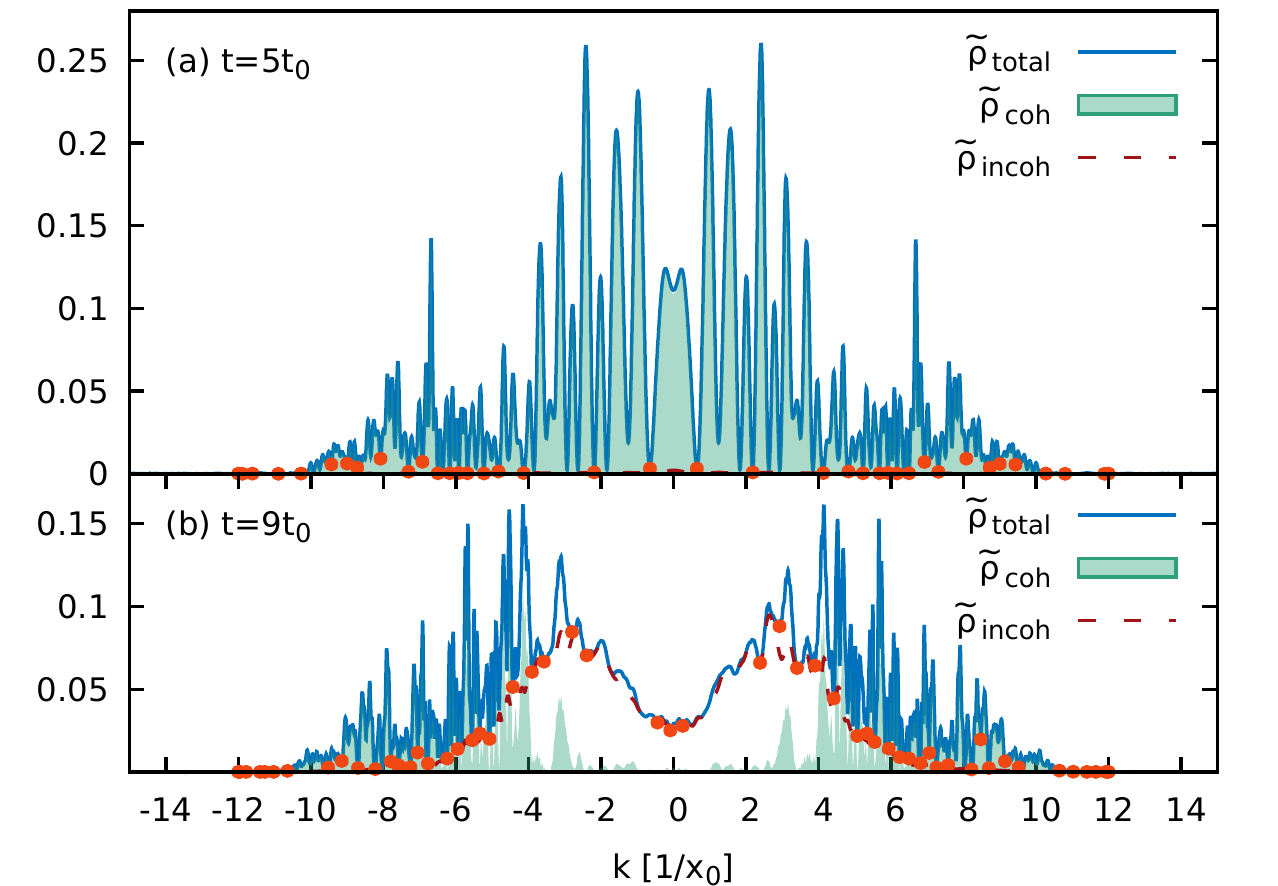}
	\caption{Total density $\tilde\rho_\text{total}(k,t)$ (solid), as well as coherent $\tilde\rho_\text{coh}(k,t)$ (filled), and incoherent part $\tilde\rho_\text{incoh}(k,t)$ (dashed) for $N=1.2\times10^6$ at (a) $t=5t_0$ and (b) $t=9t_0$. The orange dots mark the lower envelope of the strong fluctuations of $\tilde\rho_\text{total}(k,t)$, and can be used as an accurate estimate for $\tilde\rho_\text{incoh}(k,t)$. Same potential as in Fig.~\ref{fig:N6}.}
	\label{fig:incoh_den}
\end{figure}
The observed rate of exponential growth does not depend on the specific choice of the two close initial conditions such that we can clearly associate it with a Lyapunov exponent $\lambda$. The equality between $\bar d^{(2)}(t)$ and $n_\text{incoh}(t)$ proves that, if the mean-field limit is chaotic, the fraction of incoherent particles will grow exponentially with a rate given exactly by the mean-field $\lambda$. The exponentially fast depletion is thus chaos-induced, or seen from another perspective, measures mean-field chaos. Importantly, the exponential increase happens on shorter time scales than $\tau_E$, i.e.~before effects neglected within the TWA start to play a role.\\
While Fig.~\ref{fig:N6} depicts the exponential increase for $N=1.2\times10^6$ particles, we see the same exponential increase, i.e.~the same $\lambda$, also for smaller particle numbers, see Fig.~\ref{fig:N_numb}. We have varied $N$ while keeping the nonlinearity $g\propto a_sN$ constant, which amounts to increasing the scattering length $a_s$ by the same factor $N$ is decreased, which preserves the classical phase space. Indeed, with decreasing $N$ and increasing $a_s$, the BEC naturally shows larger initial depletion, but upon expansion in the periodic potential, the same $\lambda$ emerges.\\ 
We now turn to the question of how the present chaos-induced depletion could be observed in an experiment. We analyze the total particle density in momentum space $\tilde\rho_\text{total}(k,t)$ which is accessible in experiments through time-of-flight measurements, see e.g.~\cite{gericke_high-resolution_2008, erne_universal_2018}. During the expansion of the BEC, matter waves start to scatter at the potential landscape preserving initially their phase coherence. This scattering creates fluctuations in momentum space with increasingly higher frequencies as waves originating from points increasingly farther apart in real space coherently interfere. Ultimately, the density exhibits strong fluctuations reaching down to almost zero density, provided that inelastic scattering has been negligible up until this point in time, Fig.~\ref{fig:incoh_den} (a). During inelastic scattering particles lose energy, phase information, and with it, the ability to create interference fringes. These particles constitute the incoherent part of the density which piles up in form of an almost non-fluctuating background. Using a simple algorithm that determines the lower envelope of the fluctuations in the total density, we obtain a functional form very close to $\tilde\rho_\text{incoh}(k,t)$, see Fig.~\ref{fig:incoh_den} (b). Interpolating between the points of the lower envelope and integrating, we obtain $n^\text{total}_\text{low-env}$, which follows $n_\text{incoh}$ closely, see Fig.~\ref{fig:N_numb} (b). We emphasize that $n^\text{total}_\text{low-env}$ is extracted from the total density only. From Fig.~\ref{fig:N_numb} (b) it is obvious that the extraction mechanism will work best for high particle numbers with a small scattering length (e.g., for $N=1.2\times 10^6$ two orders of magnitude of exponential growth can be resolved). For smaller $N$ and correspondingly larger scattering lengths $a_s$ the incoherent density starts to pile up before coherent scattering produces sufficiently strong fluctuations in the coherent part of the density. Therefore, the close association of a non-fluctuating density with $\tilde\rho_\text{incoh}(k,t)$ is broken initially. It becomes, however, more and more accurate over time such that, in the experiment, one could observe the behavior of $\tilde\rho_\text{incoh}(k,t)$ also beyond $\tau_E$, where interferences of many-body trajectories not included within the TWA become relevant.\\
In order to measure the incoherent fraction of the total density in an experiment it is pivotal to resolve the deep minima of the fluctuations. Half of the distance between two minima is $\Delta k\gtrsim 0.04 x_0^{-1}$. Assuming a linear pixel size of a CCD camera of $2\mu$m the fluctuations could be resolved after about $300$ms time of flight. The peak amplitude of the fluctuations is $\tilde\rho_\text{total}(k)\gtrsim0.03 x_0$ leading to $\tilde\rho_\text{total}(k)\Delta k = 1.2\times10^{-3}$ such that the number of particles within each hump is greater than $100$ for $N=1.2\times10^5$ and $N=1.2\times10^6$. Despite the large time of flight necessary, we believe that the here proposed extraction could be realized in state-of-the-art BEC experiments.\\
For the disorder potential, we mostly see the same behavior as for the periodic potential, see Fig.~\ref{fig:N_numb} (c) and (d): 
\begin{figure}[t]
	\includegraphics[width=\columnwidth]{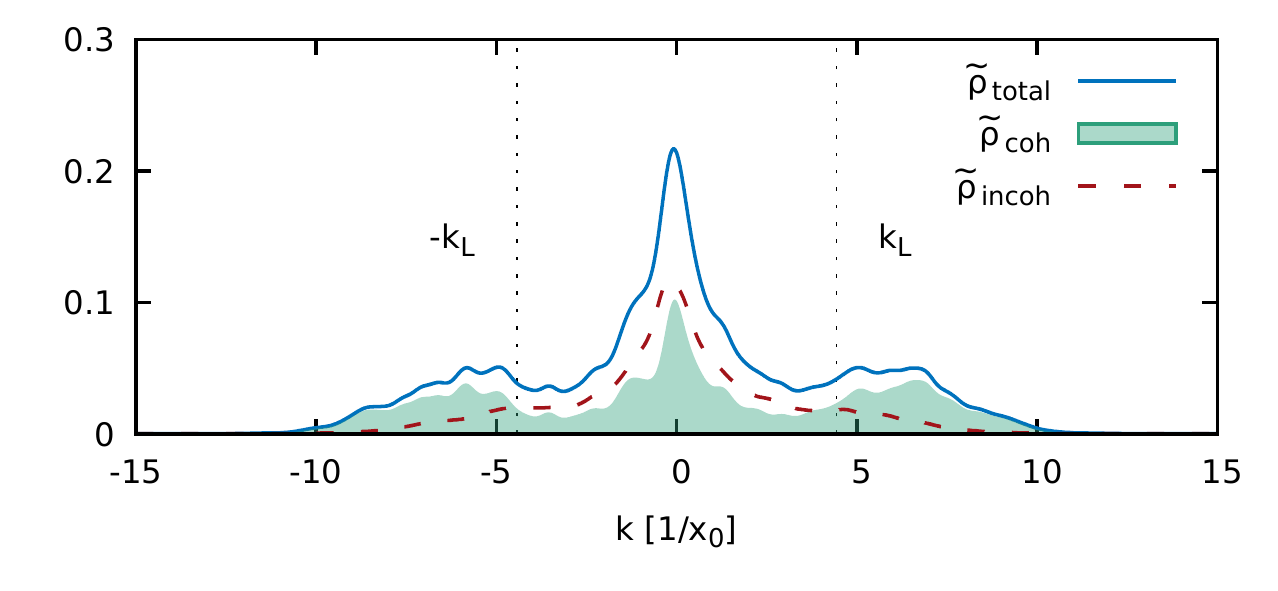}
	\caption{Particle density in momentum space at $t=8t_0$ for the speckle disorder with $V_D=0.3e$ and $\sigma =0.57\xi$ averaged over $10$ realizations (additional smoothing of the curves has been applied). The vertical lines mark the Landau velocity approximated by $k_L=\sqrt{\mu(t)}$ with $ \mu(t)$ the chemical potential at time $t$.}
	\label{fig:incoh_den_dis}
\end{figure}
$n_\text{incoh}$ grows exponentially with $\lambda$ independent of the particle number $N$. Note that we did not perform any averages over disorder realizations here. As to the extraction of the incoherent part of the density from $\tilde \rho_\text{total}(k,t)$ there is one point worth mentioning. Due to the broad spectrum of frequencies the speckle disorder offers, within few time steps, slow particles start to be scattered coherently and intertwine with particles that have lost their coherence through inelastic (i.e.~incoherent) scattering near $k=0$. The result is a local maximum in $\tilde \rho_\text{total}(k,t)$ near $k=0$, and local minima near the Landau velocity $\pm k_L$ due to inelastic scattering out of this momentum, see Fig.~\ref{fig:incoh_den_dis}. Since, however, slow particles scatter from positions in space close to each other, this scattering produces fluctuations with low frequencies as compared to the fluctuations observed for larger $k$. It is, therefore, impossible to identify $\tilde \rho_\text{incoh}(k,t)$ near $k=0$ based on the fluctuations of the total density, initially. At later times the local maximum near $k=0$ consists of incoherent particles only such that $n^\text{total}_\text{low-env}$ again accurately predicts the value of $n_\text{incoh}$, see Fig.~\ref{fig:N_numb}. For $N=1.2\times10^6$ the agreement between $n^\text{total}_\text{low-env}$ and $n_\text{incoh}$ is accurate only after $t\gtrsim12t_0$ such that we refrained from plotting it.\\
In conclusion, we have shown that a BEC expanding in a shallow periodic or disordered potential is subject to an exponentially growing depletion, and that the depletion is characterized by the ``classical" (mean-field) Lyapunov exponent $\lambda$. We have thus found a new observable that allows to identify the finger-print of classical chaos on the non-equilibrium many-body dynamics of a quantum system with a finite number of particles. In addition, we have shown how our results could be measured in an experiment by analyzing the visibility of the fluctuations of the particle density after time of flight. This opens up the possibility to verify our predictions experimentally for a real many-body system.\\

%%%%%%%%%%%%%%%%%%%%%%%%%%%%%%%%%%%%%%%%%%%%%%%%%%%%%%%%%%%%%%%%%%%%%%%%%%
%\section*{Acknowledgements}
%%%%%%%%%%%%%%%%%%%%%%%%%%%%%%%%%%%%%%%%%%%%%%%%%%%%%%%%%%%%%%%%%%%%%%%%%%
We thank Joachim Burgdörfer, David Guéry-Odelin, Dana Orsolits, Thorsten Schumm, and Juan-Diego Urbina for helpful discussions. This work has been supported by the WWTF grant MA14-002. S. D. acknowledges support by the International Max Plank Research School of Advanced Photon Science (IMPRS-APS). Calculations
were performed on the Vienna Scientific Cluster (VSC3). 

\end{document}